\documentclass[10pt,
               aps,
               pre,
               twocolumn,
               showpacs, 
               groupedaddress]{revtex4-2}
\usepackage[utf8]{inputenc}
\usepackage{graphicx}

\usepackage[dvipsnames,table]{xcolor}
\usepackage[caption=false]{subfig}
\usepackage{amsmath, amsfonts, amssymb}
\usepackage{hyperref}
\usepackage{tikz}
\graphicspath{{images/}}

\makeatletter
\newcommand*{\rom}[1]{\expandafter\romannumeral #1}
\makeatother

\begin{document}

\title{Active magneto gyrator: Memory induced trapped diamagnetism}
\author{M Muhsin}
\affiliation{Department of Physics, University of Kerala, Kariavattom, Thiruvananthapuram-$695581$, India}

\author{F Adersh}
\affiliation{Department of Physics, University of Kerala, Kariavattom, Thiruvananthapuram-$695581$, India}

\author{Mamata Sahoo}
\email{jolly.iopb@gmail.com}
\affiliation{Department of Physics, University of Kerala, Kariavattom, Thiruvananthapuram-$695581$, India}

\date{\today}

\begin{abstract}
We analytically explore the dynamics of a charged active particle coupled to two thermal baths kept at two different temperatures in two dimensions. The particle is confined to an asymmetric harmonic potential and a magnetic field of constant magnitude is applied perpendicular to the plane of motion of the particle. For such a system, as opposed to Brownian gyrator, the potential asymmetry and temperature gradient are not the key factors for the gyration, as long as finite activity and magnetic field are present. The system shows only a paramagnetic behavior in the absence of either potential asymmetry or temperature gradient. However, by tuning the temperature gradient or potential asymmetry, the system as a function of the duration of activity can exhibit paramagnetic, diamagnetic, or co-existence of both the phases. Interestingly, the magnetic moment vanishes for parameters for which the system possesses a non-equilibrium steady state and hence, a magnetic transition is observed through these non-magnetic points. Further, when the system is suspended in a viscoelastic medium characterized by a finite memory, 
it exhibits a magnetic transition in the activity-memory parameter space through a non-magnetic line.
This non-magnetic line is sensitive to temperature gradient and potential asymmetry.
It interestingly forms a closed loop with a diamagnetic phase inside the loop and the entire regime outside as paramagnetic. 
This results in the emergence of a trapped diamagnetic phase existing only within a finite regime of activity-memory parameter space. This phase eventually disappears as the temperature gradient increases (or decreases) depending on the sign of the potential asymmetry.
Moreover, it is observed that by tuning the system parameters, one can obtain zero magnetic moment even for parameter ranges that defy the equilibrium condition of the system.
\end{abstract}

\maketitle

\section{INTRODUCTION}
Active matter is an intriguing area of research due to its unique and novel features compared to that of passive matter~\cite{ramaswamy2017active,bechinger2016active, Gompper2020roadmap,pietzonka2021oddity,magistris2015intro}.
Examples of such systems include biological entities like cells~\cite{henkes2020dense}, bacterial colonies~\cite{aranson2022bacterial}, flock of birds~\cite{ballerini2008interaction}, synthetic materials like self-propelled colloids~\cite{mallory2018active}, active robots~\cite{deblais2018boundaries, dauchot2019dynamics}, and so on. 
Active matter exhibits fascinating phenomena such as spontaneous pattern formation~\cite{romanczuk2012active, martin2018collective}, phase separation~\cite{buttinoni2013dynamical, cates2015motility, jeremy2023intro, fily2012athermal}, active pressure~\cite{solon2015pressure, solon2015pressure1, takatori2014swim}, accumulation near boundaries~\cite{angelani2024optimal, marini2015towards}, anomalous diffusion~\cite{sprenger2021time, joo2020anomalous}, etc. 
The exploration of active matter provides useful insights into the fundamental principles governing non-equilibrium systems and opens avenues for designing novel materials and understanding complex biological processes~\cite{lehle2018analyzing, caprini2019active, dabelow2019irreversibility, berthier2019glassy, walther2013janus, howse2007self, palacci2013living, wang2021emergent, aguilar2016review, arijit2020active}. 
Moreover, the viscoelasticity in the environment of active particles introduces a time-dependent memory effect
that can significantly alter the behavior of active particles
~\cite{ferrer2021fluid, das2023enhanced, raikher2013brownian, grimm2011brownian}. 
For example, the viscoelasticity of the medium can enhance rotational diffusion enabling the control of the rotational motion of an active colloid~\cite{qi2020enhanced}.
It triggers fast transitions between stable states as seen in the case of a Brownian particle in a double well optical potential~\cite{ferrer2021fluid}. Similarly, a Brownian particle can also exhibit a transient caging effect in the presence of a viscoelastic medium~\cite{raikher2013brownian}.
These facts highlight the importance of considering the viscoelastic environment, as they can lead to emergent behaviors that are not present in simpler Newtonian fluids.

When a Brownian particle is coupled to two thermal baths with different temperatures along two orthogonal axes, heat usually flows from the hotter bath to the colder one, driving the system out of equilibrium~\cite{filliger2007brownian}. 
This type of physical systems characterized by two distinct temperatures find extensive applications in non-equilibrium statistical mechanics~\cite{murashita2016overdamped, nascimento2020memory, dotsenko2013two, broeck2004micro, visco2006work}. 
Such systems are fascinating because the particle can exhibit gyrating motion making them central to the study of Brownian heat engines and heat pumps~\cite{reimann2002brownian, bae2021inertial, van2006brownian, abdoli2022tunable, broek2008chiral, bustamante2005non, feldmann2000performance}.
Although the gyrating systems with two different temperatures remain mainly of theoretical interest, possible experimental realizations have been suggested in cold atoms~\cite{mancois2018two}. 
A system consisting of a Brownian particle can settle into a non-equilibrium steady state with space dependent particle currents in the presence of an anisotropic potential~\cite{dotsenko2013two}.   
Applying an external magnetic field to such a system introduces an additional layer of complexity enabling spatially correlated diffusion~\cite{abdoli2020correlations}. 
Moreover, this magnetic field can also finely tune the properties of heat engines offering a degree of control over their performance~\cite{abdoli2022tunable}. Notably, when an active particle is subjected to an external magnetic field, it can develop a non-zero orbital magnetic moment~\cite{muhsin2021orbital}, which opens up intriguing possibilities for studying gyrating systems from a new perspective. In addition, the presence of a viscoelastic environment can trigger magnetic transitions in such systems~\cite{kumar2012classical, muhsin2021orbital}.
Exploring the interplay between the gyrating motion induced by thermal asymmetry and the effects of an external magnetic field on active dynamics could yield deeper insights into the behavior of active matter systems and the design of efficient micro scale heat engines or heat pumps. 

A magneto gyrator is a system consisting of a passive charged particle in an external magnetic field, gyrating due to the presence of temperature gradient and potential asymmetry~\cite{abdoli2022tunable}. It would be interesting to explore such systems considering the motion of an active particle.
In this work, we investigate the two dimensional (2D)  dynamics of a charged active particle in an external magnetic field, in the presence of an asymmetric potential and temperature gradient. The system can exhibit gyration as a result of the temperature gradient and potential asymmetry. Additionally, it can undergo precession in the presence of a magnetic field.
To examine the effective gyrating behavior of the system, we analytically calculate the average orbital magnetic moment. 
Remarkably, the system possesses a non-zero magnetic moment even in the absence of temperature gradient or potential asymmetry, exhibiting paramagnetic behavior across the entire range of activity. 
However, with the introduction of a temperature gradient, the magnetic moment value shifts either in the positive or negative directions, leading to the emergence of a diamagnetic, paramagnetic or co-existence of both paramagnetic and diamagnetic phases. 
Interestingly, the system becomes non-magnetic for specific parameter values that even violate the equilibrium conditions. Moreover, when the particle is suspended in a viscoelastic medium characterized by a finite memory, captivatingly, we observe the emergence of a trapped diamagnetic phase surrounded by a paramagnetic regime of activity-memory parameter space. This phase gradually disappears as the temperature gradient increases or decreases depending on the sign of the potential asymmetry. In the next section, we introduce the model under consideration followed by a discussion of the main results, and finally we conclude.

\section{MODEL AND METHOD}\label{sec:model}
We consider an inertial charged self-propelling particle characterized by its mass $m$ and charge $|q|$ moving in the $x-y$ plane. The system is confined in an asymmetric potential $V$ which has the form
\begin{equation}
    V(x, y) = \frac{1}{2}k(x^2 + y^2) + \alpha x y ,
    \label{eq:potential}
\end{equation}
\begin{figure}[!hb]
    \centering
    \includegraphics[width=\linewidth]{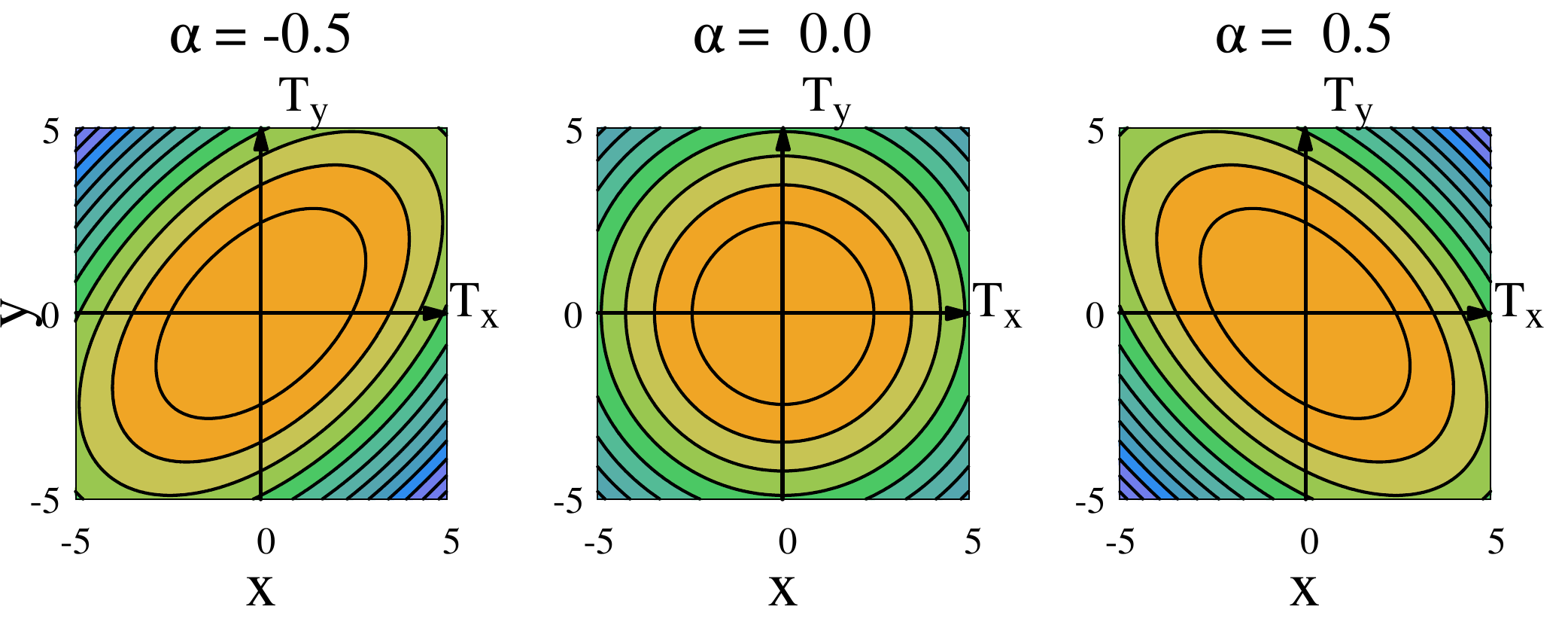}
    \caption{The 2D contour plot of the potential $V(x,y)$ [Eq.~\eqref{eq:potential}] for three different values of $\alpha$ ($-0.5$, $0$, and $0.5$) and for a fixed $k = 1$.}
    \label{fig:potential}
\end{figure}
with $\alpha$ and $k$ being the asymmetry parameter and harmonic strength of the potential, respectively. The contour plot of Eq.~\eqref{eq:potential} is shown in Fig.~\ref{fig:potential} for three different values of $\alpha$. The potential $V$ becomes harmonic when $\alpha = 0$.
An external magnetic field ${\bf B} = B {\bf \hat{k}}$ of constant amplitude $B$ acts in the $z$-direction normal to the plane of motion of the particle. 
Further, the particle is placed in a viscoelastic medium characterized by a time-dependent memory kernel in addition to a delta function kernel~\cite{sevilla2018nonequil,das2023enhanced}. 
The position vector ${\bf r} = x {\bf \hat{i}} + y {\bf \hat{j}}$ evolves according to the generalized Langevin model~\cite{igor2012visco, sevilla2019generalized} given by
\begin{equation}
\begin{split}
m \ddot{{\bf r}}(t) =& -\int\limits_{-\infty}^{t} f(t-t') \dot{{\bf r}}(t')dt' - \boldsymbol{\nabla} V + |q| (\dot{\bf r} \times {\bf B}) \\
&+ {\boldsymbol{\xi}}(t) + \boldsymbol{\eta}(t).
\end{split}
\label{eq:model}
\end{equation}

The first term of Eq.~\eqref{eq:model} represents the viscoelastic drag with $f(t - t')$ being the friction kernel. The friction kernel contains a delta function and an exponentially decaying term~\cite{das2023enhanced, sevilla2019generalized, fa2008study, raikher2013brownian}:
\begin{equation}
f\left(t-t'\right)=
\frac{\gamma}{2} \delta(t - t') + \frac{\gamma}{2t_c'}e^{-\frac{t - t'}{t_c'}},
\label{eq:kernel}
\end{equation}
with $\gamma$ being the friction or viscous coefficient of the medium. The noise $\boldsymbol{\eta}(t) = \eta_x(t) \hat{i} + \eta_y(t) \hat{j}$ is a thermal noise which has the properties $\langle\eta_i(t)\rangle = 0$ and $\langle \eta_i(t) \eta_j(t') \rangle = \delta_{ij}k_B T_i f(t - t')$ for $i,j \in \{x,y\}$.
The stochastic force $\boldsymbol{\xi}(t) = \xi_x{\bf \hat{i}} + \xi_y {\bf \hat{j}}$ appeared in Eq.~\eqref{eq:model} models the self-propulsion in the system and it follows an Ornstein-Uhlenbeck process:
\begin{equation}
t_c \dot{\boldsymbol{\xi}} = -\boldsymbol{\xi} + \xi_0\sqrt{2 t_c} \boldsymbol{\zeta}(t)
\label{eq:noise}
\end{equation}
Here, $\xi_0$ is the strength of the self-propulsion force and $\boldsymbol{\zeta}(t)$ is a delta-correlated white noise vector. $\xi_i(t)$ is of zero mean and exponentially correlated
\begin{equation}
    \langle \xi_i(t) \rangle = 0; \qquad \langle \xi_i(t) \xi_j(t') \rangle = \delta_{ij} \xi_0^2 \exp{\left( \frac{|t - t'|}{t_c} \right)}.
    \label{eq:noise_stat}
\end{equation}

Throughout the rest of this paper, the Boltzmann constant $k_B$ is considered as unity. We introduce the parameters $\Gamma = \frac{\gamma}{m}$, $\omega_0 = \sqrt{\frac{k}{m}}$, $\omega_c = \frac{|q|B}{m}$, and the temperature gradient $\Delta T = T_x - T_y$. The $\Delta T = 0$ corresponds to the same temperature of both heat baths. The average steady-state orbital magnetic moment $\langle M \rangle$ of the particle can be obtained as
\begin{equation}
    \langle M \rangle = \lim_{t \to \infty} \frac{|q|}{2} |\langle {\bf r} \times {\bf v} \rangle|,
    \label{eq:M_def}
\end{equation}
with ${\bf v} [= \dot{\bf r}]$ being the velocity of the particle.

\section{RESULTS AND DISCUSSION}\label{sec:result}
\subsection{Active Viscous Magneto Gyrator}
First, we consider the case of a viscous magneto gyrator. By taking the limit $t_c' \to 0$ on the friction kernel [Eq.~\eqref{eq:kernel}], which makes $f(t - t') = \gamma\delta(t-t')$, the system reduces to the case of a viscous magneto gyrator. In this limit, the dynamics Eq.~\eqref{eq:model} becomes
\begin{equation}
\ddot{\bf r}(t) = -\Gamma \dot{\bf r}(t) - \frac{1}{m} \boldsymbol{\nabla} V + \omega_c (\dot{\bf r} \times {\bf \hat{k}}) + \frac{1}{m} \boldsymbol{\xi}(t) + \frac{1}{m}\boldsymbol{\eta}(t).
\label{eq:model_viscous}
\end{equation}
Here, ${\bf T} = T_x \hat{i} + T_y \hat{j}$ is the temperature vector with $T_x$ and $T_y$ being the temperatures of two different baths along $x$ and $y$ axis. Using the correlation matrix method as discussed in the appendix~\ref{sec:app_A}, $\langle M \rangle$ in the steady state can be calculated and is given by
\begin{equation}
\begin{split}
\langle M \rangle &= \frac{-|q| \Delta T \alpha \Gamma}{2\left[\alpha^2 + m^2 \omega_0^2 (\Gamma^2  + \omega_c^2) \right]} \\ 
& + \frac{|q| t_c^3 \xi_0^2 \omega_c}{\Gamma \left[ -t_c^4 \alpha^2 + m^2 \left( \left( 1 + t_c\Gamma + t_c^2\omega_0^2 \right)^2 + t_c^2\omega_c^2 \right) \right]}.
\end{split}
\label{eq:M_viscous}
\end{equation}

It is to be noted that the first term of Eq.~\eqref{eq:M_viscous} is non-zero only for a finite value of $\alpha$ and $\Delta T$, i.e, as long as there is a finite asymmetry and temperature gradient in the system. Similarly, as long as a finite activity and external magnetic field are present, the second term exists and has a finite value. 
These features lead to different choices of the parameters $t_c$, $\omega_c$, $\Delta T$, and $\alpha$ for which the system exhibits different qualitative as well as quantitative behaviors. 
A non-zero value of $\omega_c$ and $t_c$ leads to the precision of the particle across magnetic field \cite{muhsin2021orbital} giving rise to a paramagnetic contribution to the system. 
On the other hand, non-zero $\alpha$ and $\Delta T$ results in the gyration of the particle either in clockwise or anticlockwise directions which can make the system either paramagnetic or diamagnetic. 
In the following sections, we discuss some of the special cases of Eq.~\eqref{eq:M_viscous} for which different qualitative behaviors are observed.

\begin{center}
    {\underline{Case 1 : Pure precession ($\alpha=0$)}}
\end{center}

The case $\alpha = 0$ corresponds to a symmetric harmonic potential. In this case, the expression of  $\langle M \rangle$ [Eq.~\eqref{eq:M_viscous}] reduces to
\begin{equation}
\langle M \rangle = \frac{|q| t_c^3 \xi_0^2 \omega_c}{m^2 \Gamma \left[ \left( 1 + t_c\Gamma + t_c^2\omega_0^2 \right)^2 + t_c^2\omega_c^2 \right]}.
\label{eq:M_viscous_a0}
\end{equation}

\begin{figure}
    \centering
    \includegraphics[width=\linewidth]{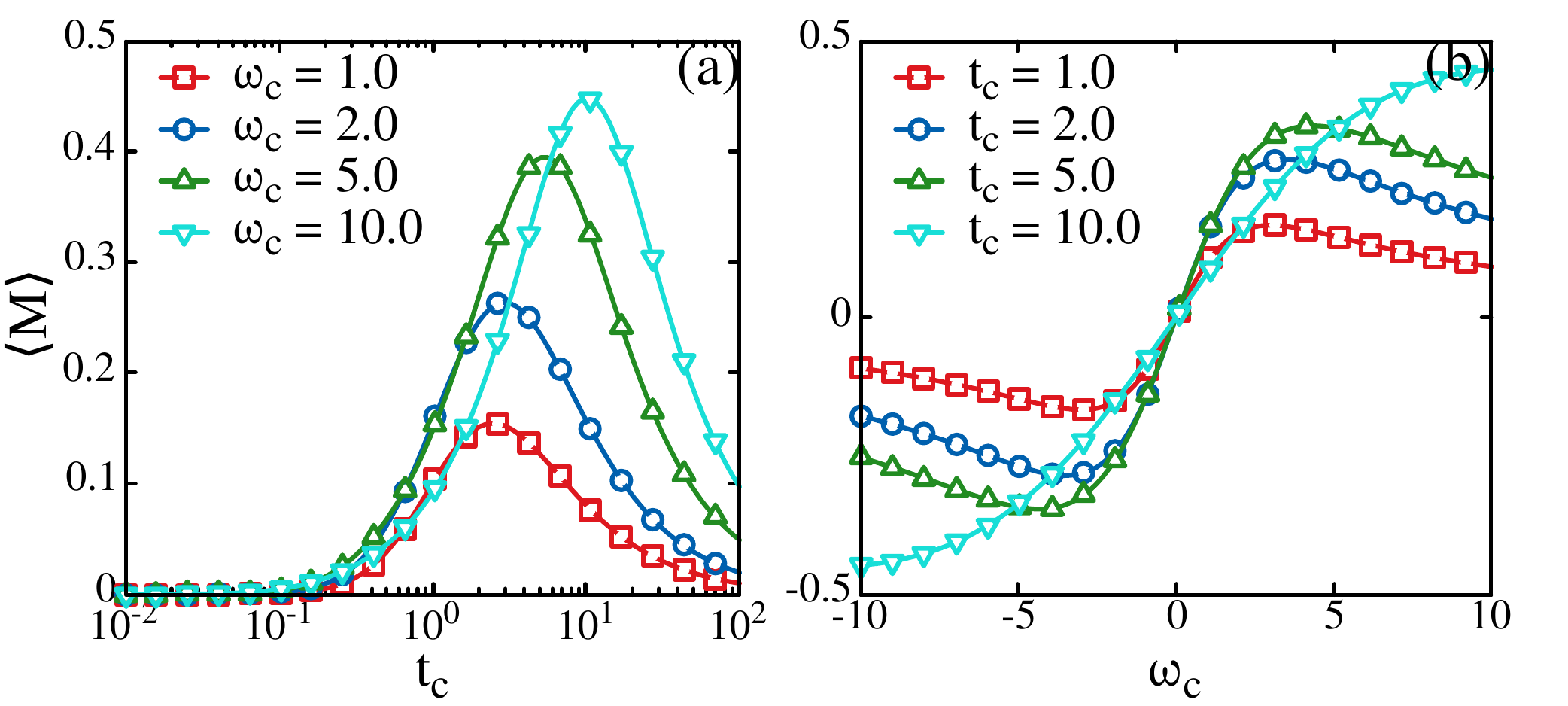}
    \caption{The $\langle M \rangle$ [Eq.~\eqref{eq:M_viscous_a0}] as a function of $t_c$ for different values of $\omega_c$ is shown in (a), as a function of $\omega_c$ for different values of $t_c$ is shown in (b). The other common parameters are : $m = \Gamma = \xi_0 = \omega_0 = q = 1$.}
    \label{fig:M_viscous_a0}
\end{figure}

Hence, the magnetic moment is zero either for $t_c = 0$ or $\omega_c = 0$. This implies that in a symmetric harmonic potential, the presence of both activity and external magnetic field is necessary in order to have a non-zero steady-state magnetic moment. 
It is evident from Eq.~\eqref{eq:M_viscous_a0} that the magnetic moment is always positive making the system always paramagnetic. In Fig.~\ref{fig:M_viscous_a0}(a), we have plotted $\langle M \rangle$ as a function of $t_c$ for different values of $\omega_c$. For a fixed $\omega_{c}$, $\langle M \rangle$ shows a nonmonotonic feature with $t_{c}$. Initially, $\langle M \rangle = 0$ in the $t_c \to 0$ limit. 
For lower values of $t_c$, we get $\langle M \rangle \approx \frac{|q|\omega_c t_c^3 \xi_0^2}{m^2 \Gamma}$ which is proportional to $t_c^3$ and hence $\langle M \rangle$ increases like $t_c^3$ and it attains a maximum as $t_c$ increases. For higher values of $t_c$, $\langle M \rangle \approx \frac{|q| \omega_c \xi_0^2}{m^2 \Gamma \omega_0^4 t_c}$ and decays to zero in $t_c \rightarrow \infty$ limit. The optimum $t_{c}$ at which $\langle M \rangle$ shows a maximum, depends on $\omega_{c}$ and shifts towards right with increase in value of $\omega_{c}$.
Similarly, $\langle M \rangle$ versus $\omega_{c}$ plots for different values of $t_{c}$ are shown in Fig.~\ref{fig:M_viscous_a0}(b).
As previously mentioned, $\langle M \rangle = 0$ for zero $\omega_c$.
With increase in $\omega_c$ value, the $\langle M \rangle$ first increases as $\langle M \rangle \approx \frac{|q|t_c^3 \xi_0^2 \omega_c}{m\Gamma \left[ 1 + t_c (\Gamma + t_c \omega_0^2) \right]^2}$ and approaches a maximum value. Further, for higher values of $\omega_c$, $\langle M \rangle$ decreases as $\langle M \rangle \approx \frac{|q| t_c \xi_0^2}{m^2 \Gamma \omega_c})$ and eventually becomes zero in $\omega_c \rightarrow \infty$ limit. The optimum $\omega_c (=\Omega_0)$ at which $\langle M \rangle$ attains its maximum is given by
\begin{equation}
    \Omega_0 = \frac{1}{t_c} + \Gamma + t_c \omega_0^2.
\end{equation}
Thus, the optimum $\omega_{c}$ ($\Omega_0$) at which $\langle M\rangle$-$\omega$ curve shows a peak depends on $t_c$, $\Gamma$, and $\omega_0$. The maximum $\langle M \rangle$ can be obtained by substituting $\omega_c = \Omega_0$ in Eq.~\eqref{eq:M_viscous_a0} and this gives
\begin{equation}
    \langle M \rangle = \frac{|q| t_c^2 \xi_0^2}{2m^2\Gamma\left[1 + t_c(\Gamma + t_c\omega_0^2)\right]}.
\end{equation}
From Eq.~\eqref{eq:M_viscous_a0}, it is to be noted that $\langle M \rangle$ is an odd function of $\omega_c$ and thus the qualitative behavior of $\langle M \rangle$ for negative values of $\omega_c$ is same as that of its positive counterpart but with the sign reversed.

\begin{center}
    \noindent{\underline{Case 2 : Pure gyration ( $\omega_c = 0$ )}}
\end{center}

When $\omega_c = 0$, the expression for the magnetic moment [Eq.~\eqref{eq:M_viscous}] becomes
\begin{equation}
    \langle M \rangle = \frac{-|q| \Delta T \alpha \Gamma}{2 (\alpha^2 + m^2 \Gamma^2\omega_0^2)},
    \label{eq:M_viscous_nomag}
\end{equation}
which is the case of a normal Brownian gyrator~\cite{mancois2018two}.
From Eq.~\eqref{eq:M_viscous_nomag}, It is confirmed that the average magnetic moment persists only for non-zero values of $\alpha$ and $\Delta T$. 
Thus, in the absence of the magnetic field, the finite magnetic moment is due to the combined effect of both temperature gradient and potential asymmetry.
For $\Delta T < 0$ ($\Delta T > 0$), the magnetic moment is positive (negative) when the asymmetry $\alpha > 0$ ($\alpha < 0$).
The mechanism of such a Brownian gyrator can be explained as follows. When $\alpha \neq 0$, due to the asymmetry of the potential, the principal axes of the potential is inclined along the left or right diagonal depending on whether $\alpha >0$ or $\alpha < 0$. The principal axis makes an angle $\frac{\pi}{4}$ with the $x$-axis when $\alpha < 0$ and $-\frac{\pi}{4}$ when $\alpha > 0$.
Thus, when $\alpha < 0$ and $\Delta T > 0$ (i.e, for $T_x > T_y$), the particle flows from hotter bath (along $x$ axis) towards the cooler bath (along $y$ axis).
This results in a net counter-clockwise gyration of the particle, resulting in a positive $\langle M \rangle$.
Similarly, for $\Delta T < 0$ or $T_x < T_y$, the particle gyration is clockwise, resulting in a negative $\langle M \rangle$. This effect is reversed in the case of $\alpha > 0$. In this case, when $\Delta T > 0$, the gyration is clockwise, resulting in a negative $\langle M \rangle$, and for $\Delta T < 0$, it is counter-clockwise, resulting in positive $\langle M \rangle$.
 The sign of the magnetic moment in Eq.~\eqref{eq:M_viscous_nomag} is determined by the product $\alpha \Delta T$. When this quantity is positive, the magnetic moment is positive and negative otherwise.

\begin{center}
    \underline{Case 3 : Gyration with precession} \\ \centering \underline{($\omega_c \neq 0$, $\alpha \neq 0$ , $t_c \neq 0$ and $\Delta T \neq 0$ )}
\end{center}

The expression of $\langle M \rangle$ in Eq.~\eqref{eq:M_viscous} is the contribution from both gyration and precession effect. The first term of Eq.~\eqref{eq:M_viscous} vanishes either for $\alpha = 0$ or $\Delta T = 0$, and the second term vanishes either for $\omega_c = 0$ or $t_c = 0$.
\begin{figure}
    \centering
    \includegraphics[width=\linewidth]{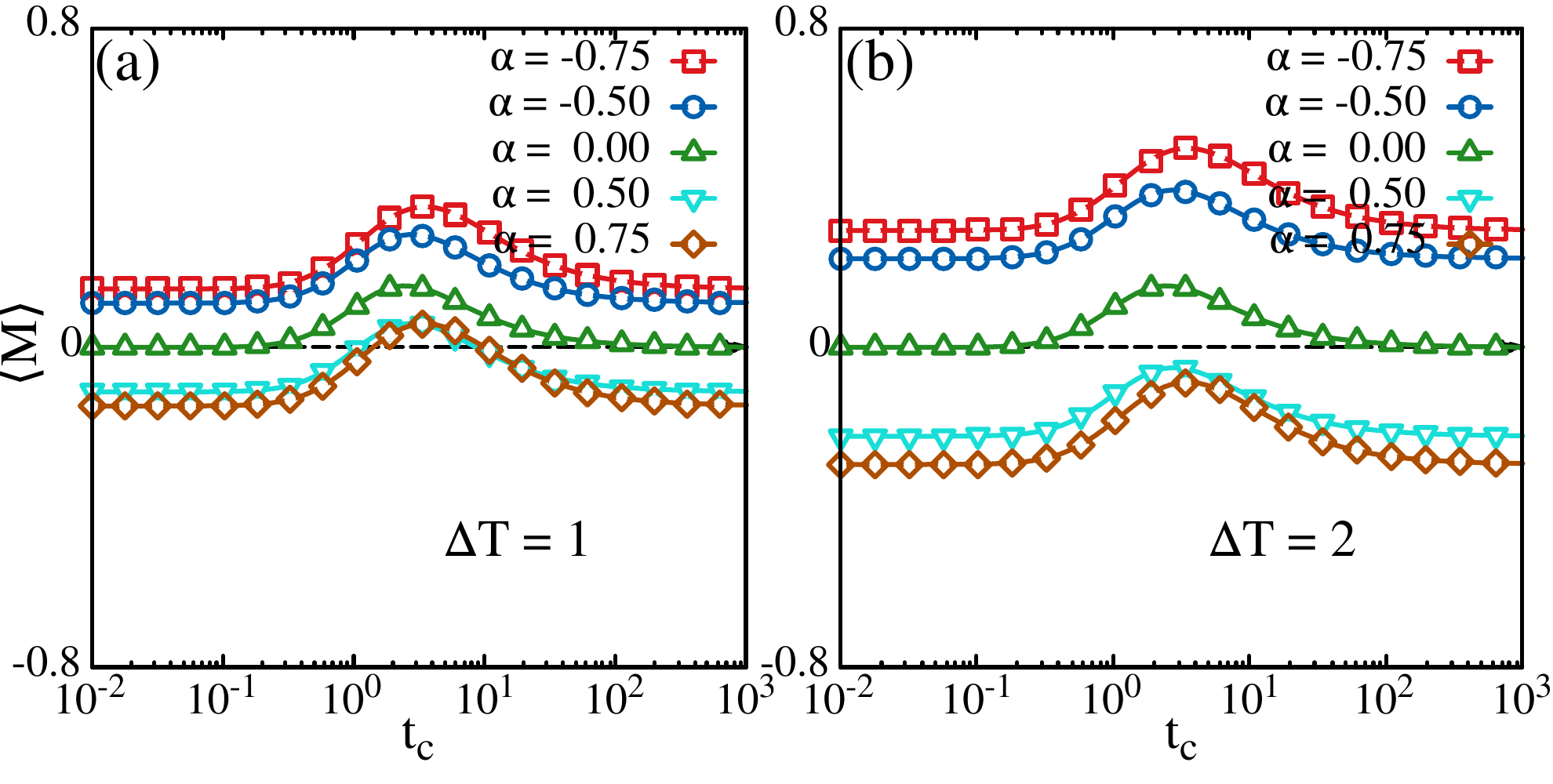}
    \caption{The $\langle M \rangle$ [Eq.~\eqref{eq:M_viscous}] as a function of $t_c$ for different values of $\alpha$ is shown in (a) for $\Delta T = 1$ and in (b) for $\Delta T = 2$. The other common parameters are : $\omega_c = \omega_0 = \Gamma = m = \xi_0 = q = 1$.}
    \label{fig:M_vs_tc_vary_a_mag}
\end{figure}
\begin{figure}
    \centering
    \includegraphics[width=\linewidth]{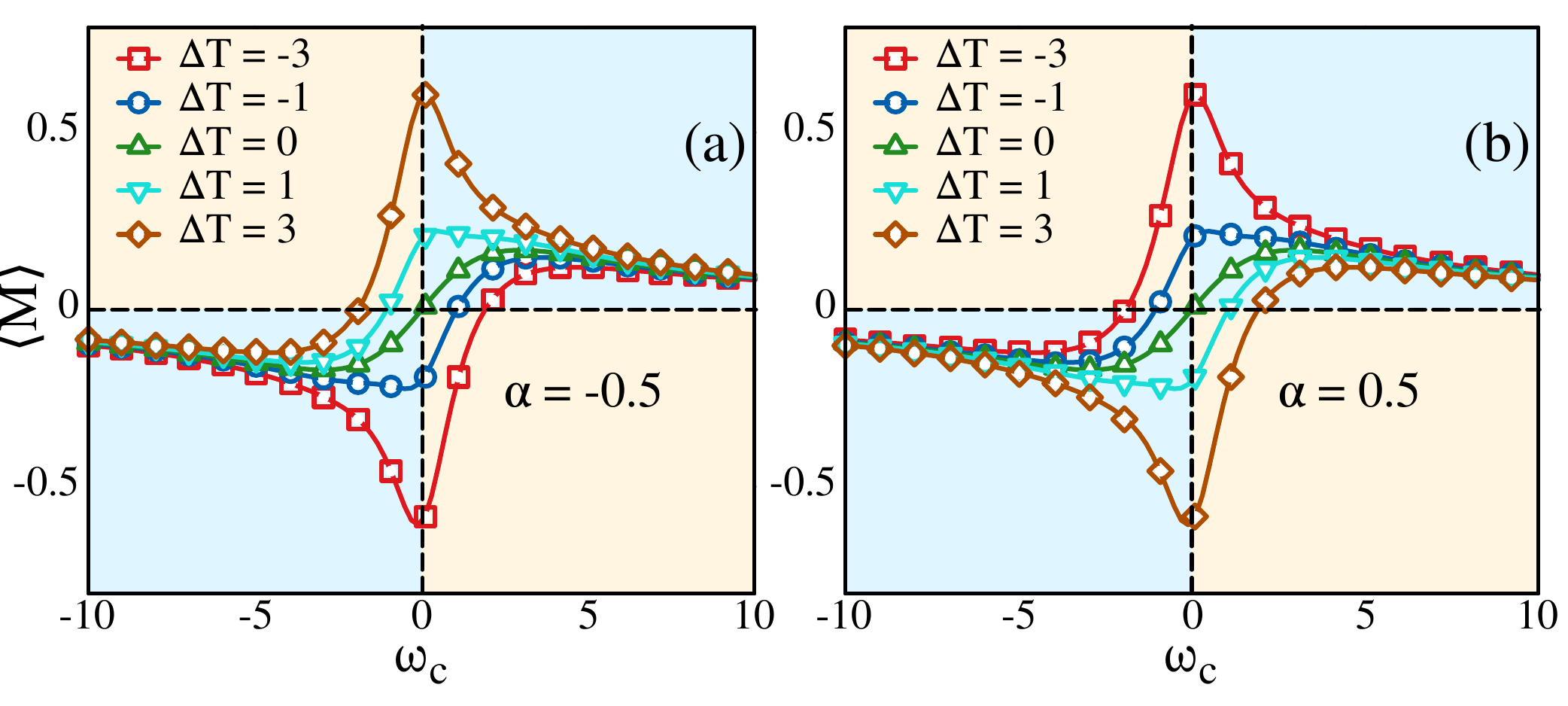}
    \caption{The $\langle M \rangle$ [Eq.~\eqref{eq:M_viscous}] as a function of $\omega_c$ for different values of $\Delta T$ in (a) for $\alpha = -0.5$ and in (b) for $\alpha = 0.5$. The other common parameters are : $t_c = \omega_0 = \Gamma = m = \xi_0 = q = 1$.}
    \label{fig:M_vs_wc_vary_Tr_mag}
\end{figure}
Figure~\ref{fig:M_vs_tc_vary_a_mag} shows the plot of $\langle M \rangle$ as a function of $t_c$ for different values of $\alpha$ and for two different values of temperature gradient $\Delta T$ [Fig.~\ref{fig:M_vs_tc_vary_a_mag}(a) and Fig.~\ref{fig:M_vs_tc_vary_a_mag} (b)]. The $\langle M \rangle$ is found to be non-monotonic as a function of $t_{c}$. In $t_c \to 0$ limit, $\langle M \rangle$ is finite and the same finite value persists for lower values of $t_{c}$. That is why $\langle M \rangle$ shows a plateau-like behavior in the lower $t_c$ regimes. In $t_{c} \rightarrow 0$ limit, $\langle M \rangle$ can be obtained as
\begin{equation}
    \langle M \rangle = \frac{-q \Delta T \alpha \Gamma}{2\left[\alpha^2 + m^2 \omega_0^2 (\Gamma^2  + \omega_c^2) \right]},
    \label{eq:M_brown}
\end{equation}
which is same as that the case of a Brownian magneto gyrator~\cite{abdoli2022tunable}. This expression confirms that in the vanishing limit of activity, both potential asymmetry and temperature gradient play a role in gyrating the particle. The null value of either of them results in a zero orbital magnetic moment. However, the expression for magnetic moment in the vanishing activity timescale is different from that of the case of the Brownian gyrator [Eq.~\eqref{eq:M_viscous_nomag}] and $\langle M \rangle$ is inversely proportional to the square of $\omega_c$. 
Further, with increase in $t_{c}$ value, $\langle M \rangle$ increases, shows a maximum and finally it decreases with $t_{c}$ and attains the same value as in the lower $t_{c}$ limit [Eq.~\eqref{eq:M_brown}].
Thus, there exists an optimum value of $t_c$ where $\langle M \rangle$ exhibits a maximum. 
If the product $\alpha \Delta T$ is negative, the system exhibits a paramagnetic behavior with positive $\langle M \rangle$. 
On the other hand, if this product is positive, the system interestingly can possess a re-entrant transition from diamagnetic to paramagnetic and then back to diamagnetic behavior for small values of $\Delta T$ [see Fig.~\ref{fig:M_vs_tc_vary_a_mag}(a)]. As the value of $\Delta T$ increases, this behavior disappears and it gives rise to a complete diamagnetic phase [Fig.~\ref{fig:M_vs_tc_vary_a_mag}(b)]. This is due to the shifting introduced by the gyration contribution to the total magnetic moment. 

\begin{figure*}
    \centering
    \includegraphics[width=0.75\linewidth]{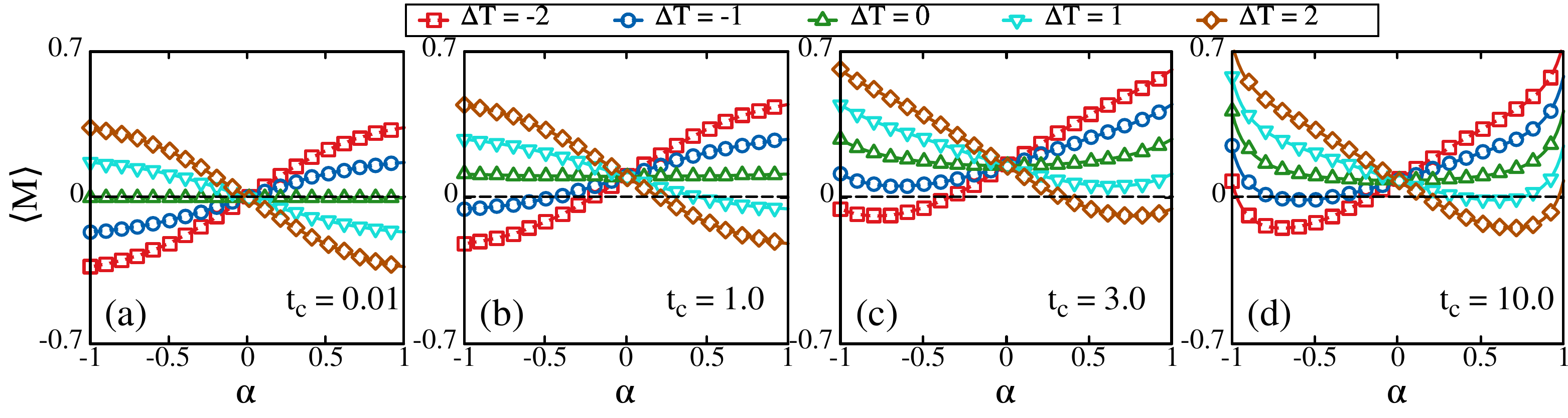}
    \caption{The plot of $\langle M \rangle$ [Eq.~\eqref{eq:M_viscous}] as a function of $\alpha$ for different values of $\Delta T$ for $t_c = 0.01$ in (a), for $t_c = 1$ in (b), for $t_c = 3$, in (c), and for $t_c = 10$ in (d). The other parameters are : $\omega_c = \Gamma = m = \omega_0 = \xi_0 = q = 1$.
    }
    \label{fig:M_vs_alpha}
\end{figure*}

\begin{figure*}
    \centering
    \includegraphics[width=0.9\linewidth]{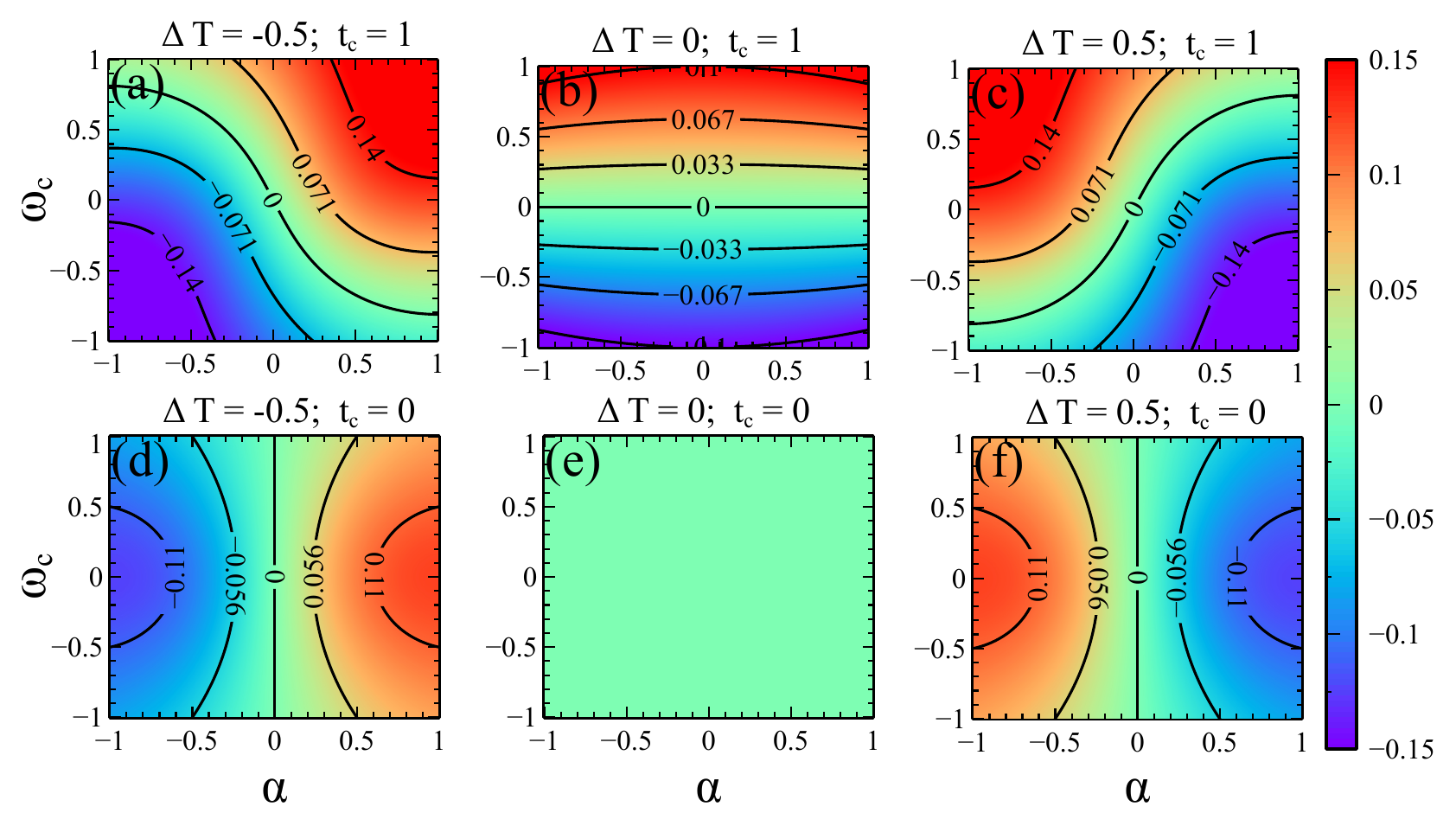}
    \caption{The 2D parametric plot of $\langle M \rangle$ [Eq.~\eqref{eq:M_viscous}] as a function of $\omega_c$ and $\alpha$ is shown in (a) and (d) for $\Delta T = -0.5$, in (b) and (e) for $\Delta T = 0$ and in (c) and (f) for $\Delta T = 0.5$. In Figs.~(a),(b) and (c), $t_c=1$ and in Figs.~(d), (e) and (f), $t_c=0$. The other common parameters are $m = \Gamma = \omega_0 = \xi_0 = q = 1$.}
    \label{fig:M_vs_wc_a}
\end{figure*}

Figure~\ref{fig:M_vs_wc_vary_Tr_mag} shows the plot of $\langle M \rangle$ as a function of $\omega_c$ for different values of $\Delta T$ and for two different $\alpha$ values in Fig.~\ref{fig:M_vs_wc_vary_Tr_mag} (a) and (b), respectively. The behavior of $\langle M \rangle$ is non-monotonic with $\langle M \rangle = 0$ for very large and small values of $\omega_c$. One can observe that the sign of $\langle M \rangle$ changes while reversing the sign of both $\omega_c$ and the product $\alpha \Delta T$ simultaneously. Thus, in addition to reversing the magnetic field, the sign of $\alpha \Delta T$ has to be changed in order for changing the sign of $\langle M \rangle$ [see Fig.~\ref{fig:M_vs_wc_vary_Tr_mag} (a) and (b)].
For $\Delta T = 0$, the system is completely paramagnetic throughout all values of $\omega_c$. When $\Delta T \neq 0$, i.e., in the presence of a temperature gradient, the system possesses a diamagnetic behavior for some ranges of $\omega_c$.

In Fig.~\ref{fig:M_vs_alpha}, we have plotted $\langle M \rangle$ as a function of $\alpha$ for different values of $\Delta T$ and for different $t_{c}$ values in Fig.~\ref{fig:M_vs_alpha}(a), Fig.~\ref{fig:M_vs_alpha} (b), Fig.~\ref{fig:M_vs_alpha}(c), and Fig.~\ref{fig:M_vs_alpha}(d), respectively. It is to be noted that for very small $t_c$ value ($t_{c}=0.01$) [see Fig.~\ref{fig:M_vs_alpha}(a)], (i.e., the case of Brownian magneto gyrator~\cite{abdoli2022tunable}), the system exhibits a dia-to-paramagnetic (for $\Delta T < 0$) and para-to-diamagnetic (for $\Delta T > 0$) transitions through $\alpha=0$ line and becomes non-magnetic with $\langle M \rangle=0$ for $\Delta T = 0$. 
This feature can also be evident from Eq.~\eqref{eq:M_viscous}. For very small $t_c$ value (i.e., $t_{c} \rightarrow 0$ limit), the contribution of the second term of Eq.~\eqref{eq:M_viscous} to the net magnetic moment is very small and hence can be ignored. Since the first term is proportional to $-\alpha \Delta T$, $\langle M \rangle$ is positive either for $\Delta T < 0$ or $\alpha < 0$ and $\langle M \rangle$ is negative only when both $\Delta T$ and $\alpha$ have the same sign. This results in the magnetic transitions as discussed above in the $\langle M \rangle$ vs $\alpha$ curves.
With further increase in $t_{c}$ value, as opposed to the case of Brownian magneto gyrator, for most of the parameter ranges, the magnetic moment shifts towards positive values [Fig.~\ref{fig:M_vs_alpha}(b) and (c)]. 
In this case, $\langle M \rangle$ is mostly positive and exhibits a magnetic transition with $\alpha$ only for $\Delta T \gg 0$ and $\Delta T \ll 0$. The system remains completely paramagnetic for the intermediate values of $\Delta T$. This behavior can be explained based on the fact that, unlike the case of Brownian magneto gyrator, for a finite $t_c$ value, there is a contribution due to the precision of the particle across the magnetic field to the total magnetic moment $\langle M \rangle$. As a result, the paramagnetic contribution to the net magnetic moment gets enhanced and hence the system becomes mostly paramagnetic for intermediate $t_{c}$ values. For higher $t_c$ values, the precession effect disappears, as a result of which $\langle M \rangle$ vs $t_c$ curve starts to shift towards negative values [Fig.~\ref{fig:M_vs_alpha}(d)].

In Fig.~\ref{fig:M_vs_wc_a}, we discuss the results from the effective contribution of both precession and gyration, by plotting a 2D plot of $\langle M \rangle$ as a function of both $\alpha$ and $\omega_c$. Fig.~\ref{fig:M_vs_wc_a}(a), (b), and (c) are plotted for three different $\Delta T$ values and for a fixed $t_c = 1$. Similarly, Figs.~\ref{fig:M_vs_wc_a}(d), (e), and (f) are plotted for three different values of $\Delta T$ and for $t_c = 0$, which is the case of a Brownian magneto gyrator. The solid dark lines of each figure correspond to the contour lines of $\langle M \rangle$. 
It can be observed that the non-magnetic line with $\langle M \rangle =0$ always passes through the origin ($\omega_c = 0, \alpha=0$) of the $\omega_{c}-\alpha$ plane. This is because, this point corresponds to the case with no asymmetry and the absence of magnetic field. The locus of all points on this non-magnetic line satisfies the equation
\begin{equation}
    \frac{t_c^3 \omega_c \left[ \alpha^2 + m^2\omega_0^2(\Gamma^2 + \omega_c^2) \right]}{\Gamma^2\left[ -t_c^4\alpha^3 + m^2\alpha \left[ \left( 1 + t_c(\Gamma + t_c\omega_0^2) \right)^2 + t_c^2 \omega_c^2 \right] \right]} = \frac{\Delta T}{2 \xi_0^2}.
   \label{eq:non-magnetic-locus}
\end{equation}
The existence of this non-magnetic line implies that for the parameter values on this line, both gyration and precession effects get canceled with each other, resulting in a net null magnetic moment. 
When $\Delta T = 0$ and for a finite $t_c$ value, [Fig.~\ref{fig:M_vs_wc_a}(b)], the equation for the non-magnetic line [Eq.~\eqref{eq:non-magnetic-locus}] reduces to $\omega_c = 0$, confirming no transition. However, the average magnetic moment $\langle M \rangle$ is positive for $\omega_c > 0$ and vice versa. This is due to the fact that $\langle M \rangle$ is an odd function of $\omega_{c}$.
On the other hand, for $\Delta T < 0$ [Fig.~\ref{fig:M_vs_wc_a}(a)], the non-magnetic line crosses the $\omega_c-\alpha$ parametric plane from the top left to the bottom right diagonal. As a result, $\langle M \rangle$ transforms from a negative value to a positive value along the bottom-left to top-right diagonal of the parametric plane, confirming a transition from the diamagnetic to paramagnetic phase.
Similarly, for $\Delta T > 0$ [Fig.~\ref{fig:M_vs_wc_a}(c)], the non-magnetic line passes from the top-right to the bottom-left diagonal of $\omega_c-\alpha$ parametric plane. In this case, a magnetic transition from diamagnetic phase to paramagnetic phase is observed from the bottom-right corner of the parametric plane to the top-left corner. 
It can be seen that for both of these above two cases, the system exhibits a transition from diamagnetic to paramagnetic phase through a non-magnetic line for which even both $\omega_{c}$ (or $t_{c}$) and $\alpha$ (or $\Delta T)$ are non-zero [see Fig.~\ref{fig:M_vs_wc_a}(a) and (c)]. Thus, the system exhibits a null magnetic moment even for the parameters for which it doesn't validate the equilibrium condition, i.e., for $\Delta T\neq 0$ and $t_{c} \neq 0$. This is because, in $t_{c} \rightarrow 0$ limit and for $\Delta T=0$, the system becomes passive and approaches thermal equilibrium.

The Figs.~\ref{fig:M_vs_wc_a}(d),~\ref{fig:M_vs_wc_a}(e), and ~\ref{fig:M_vs_wc_a}(f) represent the cases for which we consider $t_{c}$ as zero. This is same as the case of a Brownian magneto gyrator. The Fig.~\ref{fig:M_vs_wc_a}(e) corresponds to the case of a passive Brownian particle as for $\Delta T=0$ and $t_{c}=0$, the system approaches thermal equilibrium and the particle becomes passive. That is why $\langle M \rangle$ is observed to be zero throughout the entire $\omega_c - \alpha$ parametric plane [Fig.~\ref{fig:M_vs_wc_a}(e)].
However, a magnetic transition is noticed for $\Delta T \neq 0$ [Fig.~\ref{fig:M_vs_wc_a}(d) and (f)] through the non-magnetic line $\alpha = 0$. For $\Delta T \le 0$ [Fig.~\ref{fig:M_vs_wc_a}(d)], the magnetic transition is observed from diamagnetic to paramagnetic phase through $\alpha=0$ line. Similarly, for $\Delta T\ge 0$ [Fig.~\ref{fig:M_vs_wc_a}(f)], the magnetic transition is observed from the paramagnetic to diamagnetic phase through $\alpha=0$ line. 
Interestingly, for these above two cases [Fig.~\ref{fig:M_vs_wc_a}(d) and Fig.~\ref{fig:M_vs_wc_a}(f)], $\langle M \rangle$ is found to be symmetric about the $\omega_c = 0$ line. The magnitude of $\langle M \rangle$ does not change while reversing the direction of the magnetic field.

\subsection{Active Viscoelastic Magneto Gyrator}
When the particle is suspended in a viscoelastic environment, the dynamics is given by Eq.~\eqref{eq:model}. The expression of $\langle M \rangle$ in this case is given by

\begin{figure*}
    \centering
    \includegraphics[width=\linewidth]{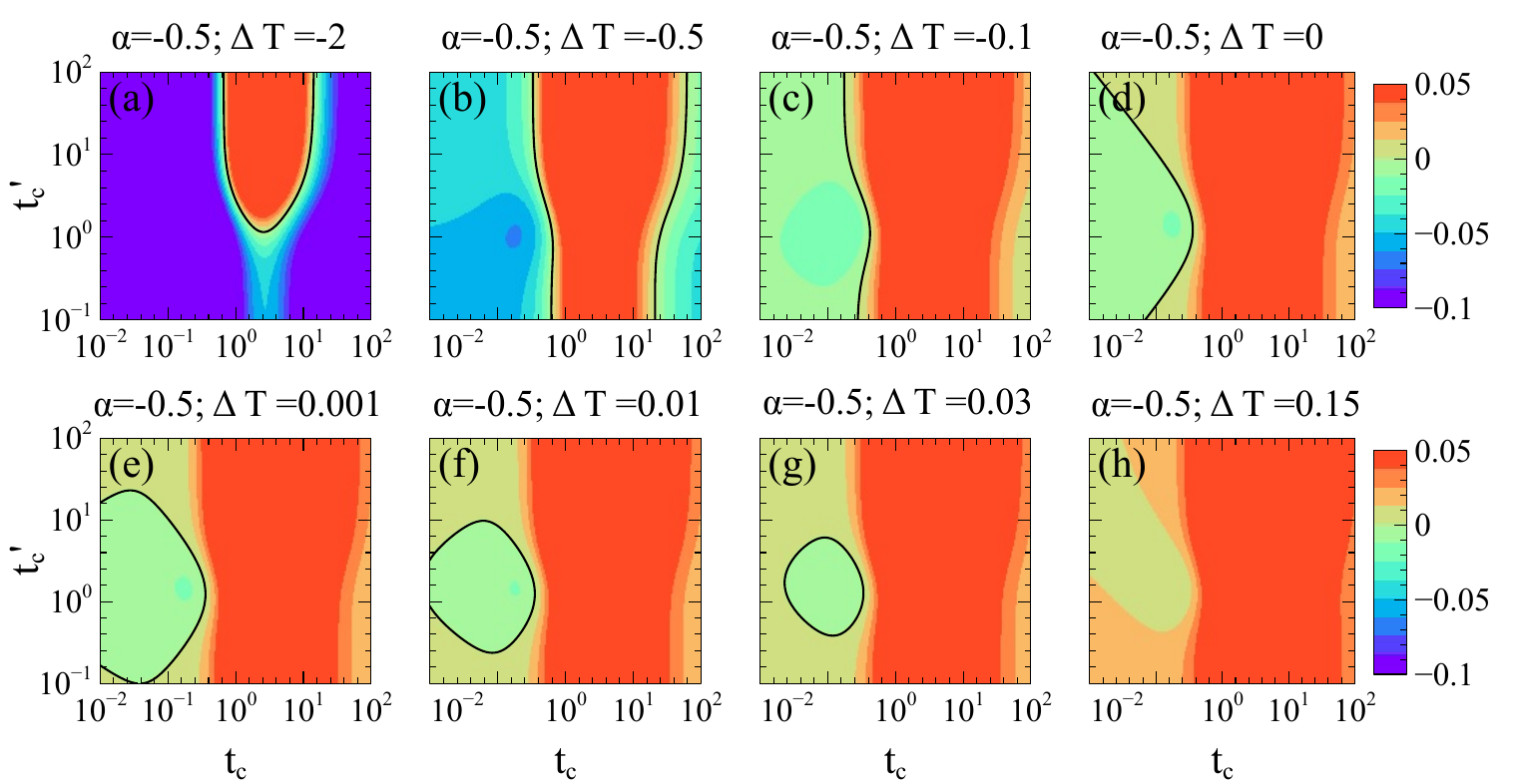}
    \caption{The 2D parametric plot of $\langle M \rangle$ [Eq.~\eqref{eq:M_gen}] as a function of $t_c$ and $t_c'$ is shown in (a) for $\Delta T = -2$, (b) for $\Delta T = -0.5$, (c) for $\Delta T = -0.1$, (d) for $\Delta T = 0$, (e) for $\Delta T = 0.001$, (f) for $\Delta T = 0.01$, (g) for $\Delta T = 0.03$, and in (h) for $\Delta T = 0.15$. The color map represents the evolution of magnetic moment. The other common parameters are $\Gamma = m = \omega_0 = \omega_c = \xi_0 = q = 1$, and $\alpha = -0.5$.}
    \label{fig:M_vs_tc_tcp}
\end{figure*}
\begin{widetext}
\begin{equation}
		\begin{split}
			\langle M\rangle = & \frac{|q| \Delta T \alpha \Gamma\left[ -2(t_c')^4\alpha^2 + (B + 2 (t_c')^2 (\omega_0^2 + \omega_c^2)) m^2 \right]}{S}\\
            & + \frac{2|q| t_c \xi_0^2 \omega_c \left[ -4t_c^2(t_c')^4 (t_c + t_c')^2 \alpha^2 + m^2 \left( -4(t_c')^4 - 4 t_c (t_c')^3 (2 + t_c' \Gamma) + A t_c^2(t_c')^2 + 4B t_c^3 t_c' + 2(B - (t_c')^3\Gamma \omega_0^2) t_c^4 \right) \right]}{S'},
		\end{split}
		\label{eq:M_gen}
\end{equation}
with the symbols $S$, $S'$, $A$, and $B$ are given by
\begin{align}  
	S =& 2 m^4 \omega_0^2 \left[-\omega_c^2 \left(t_c' \left(4 (t_c')^3 \omega_0^4+8 t_c' \omega_0^2 (\Gamma  t_c'+1)+ 5 \Gamma^2  t_c'+4\Gamma\right)+4\right)-4 (t_c')^2 \omega_c^4-\Gamma ^2 \left(t_c' \left(\Gamma +t_c' \omega_0^2\right)+2\right)^2\right] + 8 \alpha ^4 (t_c')^4 \nonumber\\
    & +\alpha ^2 m^2 \left[2 (t_c')^4 \omega_0^2 \left(\Gamma ^2-4 \omega_0^2+4 \omega_c^2\right)+\Gamma  (t_c')^3 \left(\Gamma ^2-12 \omega_0^2+4 \omega_c^2\right)+2 (t_c')^2 \left(\Gamma ^2-8 \omega_0^2-4 \omega_c^2\right)-4 \Gamma  t_c'-8\right],\\
	S' =& \Gamma  \left[m^2 \left(2 (t_c')^2 \omega_c^2+\left(t_c' \left(\Gamma +t_c' \omega_0^2\right)+2\right)^2\right)-\alpha ^2 (t_c')^4\right] \nonumber\\
        & \left[m^2 \left(4 t_c^2 \omega_c^2 (t_c+t_c')^2+\left(t_c \left(2 \Gamma  t_c+2 t_c \omega_0^2 (t_c+t_c')+\Gamma  t_c'+2\right)+2 t_c'\right)^2\right)-4 \alpha ^2 t_c^4 (t_c+t_c')^2\right],\\
	A = & t_c' \left[\Gamma  \left(8 (t_c')^2 \omega_0^2+4\right)+4 t_c' \left((t_c')^2 \omega_0^4+2 \omega_0^2+\omega_c^2\right)+3 \Gamma ^2 t_c'\right]+4, \quad \text{and}\\
	B = & 2 (t_c')^4 \omega_0^4+3 \Gamma  (t_c')^3 \omega_0^2+(t_c')^2 \left[\Gamma ^2+4 \omega_0^2+2 \omega_c^2\right]+4 \Gamma t_c'+4 .
\end{align}
\end{widetext}

In Fig.~\ref{fig:M_vs_tc_tcp}, we have shown the 2D plot of $\langle M \rangle$ as a function of both $t_c$ and $t_c'$. In $t_c$-$t_c'$ parameter space, depending on the value of $\alpha$ and $\Delta T$, complete paramagnetism, complete diamagnetism, or co-existence of both para and diamagnetic behavior is observed. 
Initially, for a negative $\alpha$ and negative $\Delta T$ [Fig.~\ref{fig:M_vs_tc_tcp}(a)], the $t_c$-$t_c'$ parameter space is dominated by the diamagnetic regime with a small paramagnetic regime present for higher values of $t_c'$ and for intermediate values of $t_c$. Both these phases co-exists via a non-magnetic line for the parameters which does not satisfy the equilibrium condition. The same can also be evident from Eq.~\eqref{eq:M_gen}, i.e., when both $\Delta T$ and $\alpha$ are negative, the first term of Eq.~\eqref{eq:M_gen} dominates over the second one, resulting $\langle M \rangle$ being negative for the most of the parameter regime. 
With further increase in $\Delta T$ towards $\Delta T = 0$, the paramagnetic regime starts to expand towards both higher and lower $t_{c}$ regimes [see Fig.~\ref{fig:M_vs_tc_tcp}(b) and (c)] and exactly for $\Delta T = 0$ [Fig.~\ref{fig:M_vs_tc_tcp}(d)], the diamagnetic phase exists only for lower $t_c$ regime. 
This is because, for $\Delta T=0$ and in the $t_c \to \infty$ limit, both gyrating and precision effects disappear, resulting in a null magnetic moment and making the system non-magnetic in the higher $t_{c}$ regimes. 
However, for $\Delta T=0$ and in the lower $t_c$ regime, the dominant diamagnetic contribution might be due to the precision effect in a direction opposite to the magnetic field (the contribution from the second term of Eq.~\eqref{eq:M_gen}), resulting a memory induced diamagnetism.
Interestingly, at this point, in contrast to that of viscous case, a diamagnetic phase exists for a finite $t_c'$, even if $\Delta T = 0$, which is evident from Eq.~\eqref{eq:M_gen}.
For $\Delta T = 0$ and taking small $t_c$ limit of Eq.~\eqref{eq:M_gen}, we get
\begin{equation}
    \langle M \rangle = -\frac{2|q|(t_c')^2 \xi_0^2 \omega_c t_c}{\Gamma D_1} + O(t_c^2),
    \label{eq:M-Dt0-small-tc}
\end{equation}
with 
\begin{equation}
    D_1 = -(t_c')^4\alpha^2 + m^2\left[\left( 2 +  t_c'\Gamma + (t_c')^2\omega_0^2 \right)^2 + 2(t_c')^2 \omega_c^2\right].
\end{equation}
It is to be noted from Eq.~\eqref{eq:M-Dt0-small-tc} that the dominant term is diamagnetic, hence the observed diamagnetic behavior at lower $t_c$ regime [Fig.~\ref{fig:M_vs_tc_tcp}(d)]. In $t_c' \to 0$ limit, this term vanishes, which implies that the memory has a crucial role in making the system diamagnetic at this point.
When $\Delta T$ is further increased towards the positive side, the diamagnetic phase further shrinks and finally becomes surrounded by paramagnetic regimes [see Figs.~\ref{fig:M_vs_tc_tcp}(f) and (g)]. As a result, the non-magnetic line forms a closed loop with a diamagnetic phase existing inside the loop. For large values of $\Delta T$, the memory-induced diamagnetic contribution gets canceled with the gyrating effect, resulting in the disappearance of the diamagnetic phase and only the paramagnetic phase exists. The same behavior can be seen when $\alpha$ is taken as positive and $\Delta T$ is decreased from a positive value towards a negative value. This is because, changing the signs of both $\alpha$ and $\Delta T$ has no effect on the expression of $\langle M \rangle$ in Eq.~\eqref{eq:M_gen}. That is, $\langle M \rangle(-\alpha, -\Delta T) = \langle M \rangle(\alpha, \Delta T)$.

\begin{figure}
    \centering
    \includegraphics[width=\linewidth]{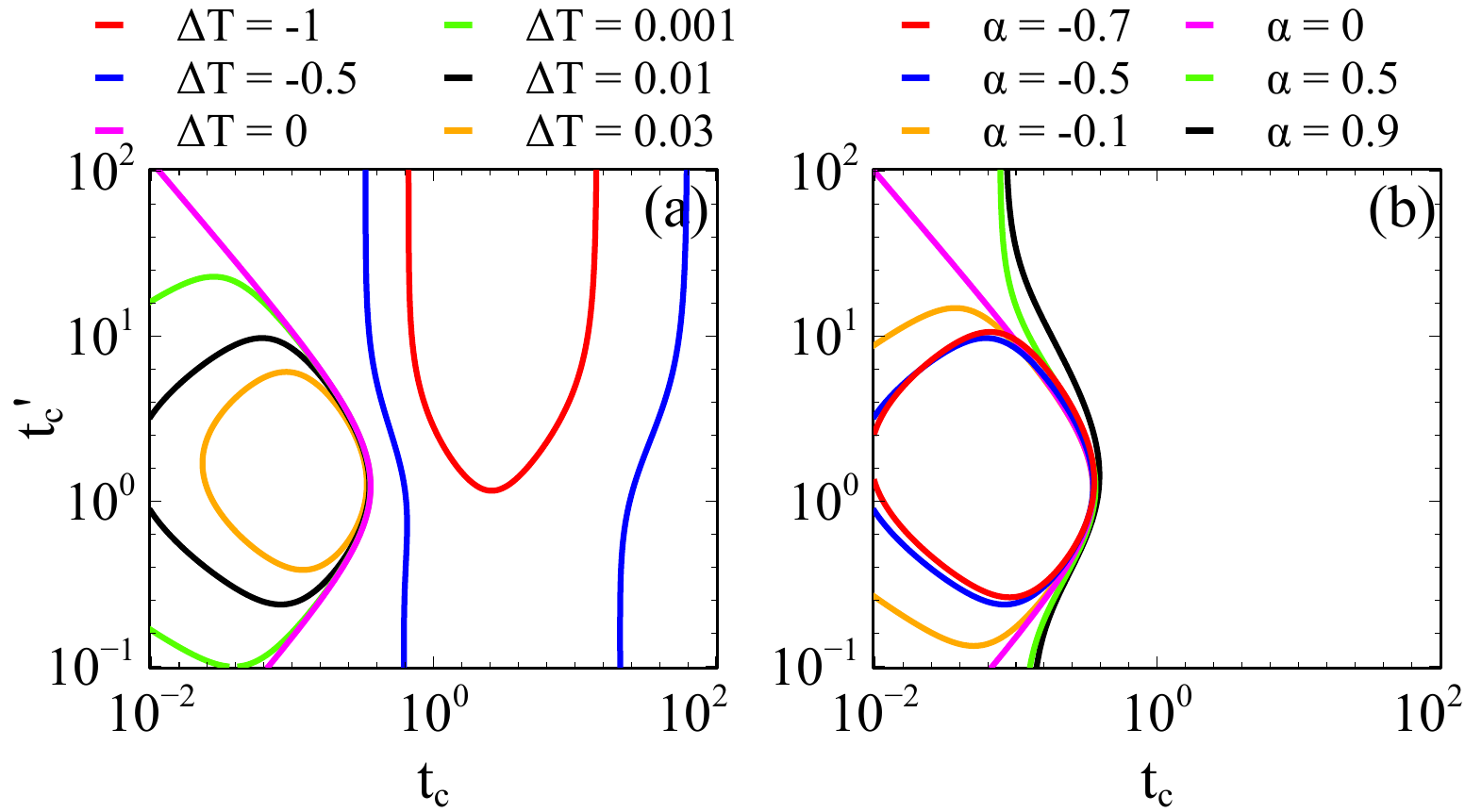}
    \caption{The 2D plot of non-magnetic lines in $t_c-t_c'$ parameter space is shown in (a) for different values of $\Delta T$ for a fixed $\alpha = -0.5$, and in (b) for different values of $\alpha$ for a fixed $\Delta T = 0.01$. The other common parameters are $\Gamma = m = \omega_0 = \omega_c = \xi_0 = q = 1$.}
    \label{fig:NM-contour}
\end{figure}
These results are further confirmed in Fig~\ref{fig:NM-contour}, where we have plotted the non-magnetic lines in $t_{c}-t_{c}'$ plane for different values of $\alpha$ and $\Delta T$ in Fig~\ref{fig:NM-contour} (a) and Fig~\ref{fig:NM-contour} (b), respectively. The non-magnetic lines are the loci of all the points for which $\langle M \rangle = 0$. 
These non-magnetic lines divide the $t_c-t_c'$ plane into the diamagnetic and paramagnetic phases. In  Fig.~\ref{fig:NM-contour}(a), $\alpha$ is considered to be $-0.5$ and $\Delta T$ is increased from a negative value across $\Delta T = 0$. 
As $\Delta T$ increases, the non-magnetic line shifts towards both higher and lower values of $t_c$ in $t_c-t_c'$ plane. The line on the higher side of $t_{c}$ disappears at $\Delta T = 0$, making the parametric plane exactly into two phases. 
Further increase in $\Delta T$ value (close to zero) results in non-magnetic line forming a closed loop. In this case, the finite region inside the loop has a diamagnetic behavior, while the region outside the loop is of paramagnetic nature. 
This closed region starts to shrink and the area inside the loop decreases as the temperature gradient is further increased and disappears for larger values of $\Delta T$. The existence of this closed loop also depends on the value of $\alpha$. 
The dependence of the non-magnetic lines on $\alpha$ is shown in Fig~\ref{fig:NM-contour}(b) for a fixed $\Delta T = 0.01$. 
The closed loop appears when $\alpha$ is negative. The area inside the loop increases as the value of $\alpha$ increases from negative value towards zero, and the loop becomes opened for $\alpha \ge 0$. 
Thus, the appearance of closed loop depends on the magnitude of $\alpha$, $\Delta T$, and the sign of the product $\alpha \Delta T$. 
When the sign of the product is negative, the non-magnetic lines become closed for small enough values of $\Delta T$.

\section{CONCLUSIONS}\label{sec:summary}
In this work, we have investigated the self-propulsion of a charged active particle in two dimensions. The particle is confined to an asymmetric harmonic potential with two different temperatures along $x$ and $y$ axes. Additionally, a magnetic field of constant magnitude is applied perpendicular to the plane of motion of the particle. At steady state, we have exactly calculated the average orbital magnetic moment of the particle using the correlation matrix method. The calculated magnetic moment has two parts. One part is due to the temperature gradient and finite asymmetry of the potential. 
The other part is due to the presence of activity and the external magnetic field. In a viscous medium, the three cases of interest are pure precession, pure gyration, and gyration with precession. 
In the case of pure precession, the magnetic moment is non-monotonic with the magnetic field and activity and the system is always paramagnetic in nature. However, in the presence of both gyration and precession, the magnetic moment value shifts either in the positive or negative directions, depending on the sign of the potential asymmetry and temperature gradient. This results in a completely paramagnetic, completely diamagnetic, or the co-existence of both paramagnetic and diamagnetic phases. Interestingly, the system exhibits a magnetic transition from diamagnetic to paramagnetic (or vice-versa) phase through a non-magnetic line in a magnetic field - potential asymmetry parameter space. This non-magnetic line passes through the loci of points for which the equilibrium condition is not validated.

Further, when the particle is subjected to a viscoelastic suspension characterized by a finite memory, the system can exhibit a diamagnetic, a paramagnetic, or co-existence of both the phases through a non-magnetic line in the parametric space spanned by memory and activity timescale.
Interestingly, when the potential asymmetry and temperature gradient are of opposite signs, the non-magnetic line forms a closed loop with a diamagnetic regime inside the loop and the entire regime outside as paramagnetic, resulting in an appearance of a trapped diamagnetic phase within a finite regime of activity-memory parameter space.
This phase gradually disappears with further increase or decrease in the temperature gradient depending on the sign of the potential asymmetry. The arrival of this engrossing feature might be attributed due to the complex interplay of both gyrating effect and the precession effect. Finally, we believe that our analysis can be amenable for experimental verification and might apply to skyrmion kind particles~\cite{mochizuki2014thermal}. Further, it is interesting to look at how such a system can be used as a heat engine or refrigerator and this will be reported elsewhere.

\section{Acknowledgement}
MS thanks Arnab Saha for some valuable discussions. MS acknowledges the start-up grant from UGC, state plan fund from the University of Kerala, SERB-SURE grant (SUR/2022/000377), and CRG grant (CRG/2023/002026) from DST, Govt. of India for financial support.

\appendix
\section{} \label{sec:app_A}
For the calculation of magnetic moment, we first express the dynamics Eq.~\eqref{eq:model} in matrix form. The thermal noise $\boldsymbol{\eta}(t)$ in Eq.~\eqref{eq:model} can be taken as the sum of two independent Gaussian noises, $\boldsymbol{\eta}(t)=\boldsymbol{\eta_1}(t) + \boldsymbol{\eta_2}(t)$. Here $\boldsymbol{\eta_1}(t)$ is a white noise with the correlation $\langle{\eta_1}_i(t){\eta_1}_j(t)\rangle~=~\delta_{ij}\gamma k_BT_i \delta(t - t')$, and $\boldsymbol{\eta_2}(t)$ is a colored noise with the correlation $\langle{\eta_2}_i(t){\eta_2}_j(t)\rangle~=~\delta_{ij}\frac{\gamma}{2t_c'}k_BT_i e^{-(t - t')/t_c'}$. The noise $\boldsymbol{\eta_2}(t)$ can be modelled as an Ornstein-Uhlenbeck process as
\begin{equation}
    t_c'\boldsymbol{\dot{\eta_2}}(t) = -\boldsymbol{\eta_2}(t) + \begin{pmatrix}
    \sqrt{\gamma k_B T_x} & 0 \\
    0 & \sqrt{\gamma k_B T_y}
    \end{pmatrix} \cdot \boldsymbol{\zeta}'(t),
\end{equation}
with $\boldsymbol{\zeta}'(t)$ a delta correlated white noise. To rewrite Eq.~\eqref{eq:model} as a matrix equation, we introduce the vector ${\bf U}$ such that
\begin{equation}
    {\bf U} = \int\limits_{-\infty}^{t} \frac{\Gamma}{2t_c'} e^{-\frac{t - t'}{t_c'}} {\bf \dot{r}}(t')\; dt' 
\end{equation}
Hence, the Eq.~\eqref{eq:model} can be expressed as
\begin{equation}
    \dot{\boldsymbol{\chi}} = A\boldsymbol{\chi} + B \boldsymbol{\eta'},
    \label{eq:dynamics-matrix}
\end{equation}
with the matrices $\boldsymbol{\chi}$, $A$, $B$, and $\boldsymbol{\eta'}$ are given by
\begin{equation}
    \boldsymbol{\chi} = (x\ y\ v_x\ v_y\ U_x\ U_y\ \xi_x\ \xi_y\ \eta_{2x}\ \eta_{2y})^T,
\end{equation}
\begin{widetext}
\begin{equation}
    A = \begin{pmatrix}
        0 & 0 & 1 & 0 & 0 & 0 & 0 & 0 & 0 & 0 \\
        0 & 0 & 0 & 1 & 0 & 0 & 0 & 0 & 0 & 0 \\
        -\omega_0^2 & -\frac{\alpha}{m} & -\frac{\Gamma}{2} & \omega_c & -1 & 0 & \frac{1}{m} & 0 & \sqrt{\frac{\Gamma k_B T_x}{m}} & 0 \\ 
        -\frac{\alpha}{m} & -\omega_0^2 & -\omega_c & -\frac{\Gamma}{2} & 0 & -1 & 0 & \frac{1}{m} & 0 & \sqrt{\frac{\Gamma k_B T_y}{m}} \\
        0 & 0 & \frac{\Gamma}{2t_c'} & 0 & \frac{-1}{t_c'} & 0 & 0 & 0 & 0 & 0 \\
        0 & 0 & 0 & \frac{\Gamma}{2t_c'} & 0 & \frac{-1}{t_c'} & 0 & 0 & 0 & 0 \\
        0 & 0 & 0 & 0 & 0 & 0 & \frac{-1}{t_c} & 0 & 0 & 0 \\
        0 & 0 & 0 & 0 & 0 & 0 & 0 & \frac{-1}{t_c} & 0 & 0 \\
        0 & 0 & 0 & 0 & 0 & 0 & 0 & 0 & \frac{-1}{t_c'} & 0 \\
        0 & 0 & 0 & 0 & 0 & 0 & 0 & 0 & 0 & \frac{-1}{t_c'}
    \end{pmatrix},
\end{equation}
\begin{equation}
    B = \begin{pmatrix}
        0 & 0 & 0 & 0 & 0 & 0 & 0 & 0 & 0 & 0 \\
        0 & 0 & 0 & 0 & 0 & 0 & 0 & 0 & 0 & 0 \\
        0 & 0 & \sqrt{\frac{\Gamma k_BT_x}{m}} & 0 & 0 & 0 & 0 & 0 & 0 & 0 \\
        0 & 0 & 0 & \sqrt{\frac{\Gamma k_BT_y}{m}} & 0 & 0 & 0 & 0 & 0 & 0 \\
        0 & 0 & 0 & 0 & 0 & 0 & \sqrt{\frac{2 \xi_0^2}{t_c}} & 0 & 0 & 0 \\
        0 & 0 & 0 & 0 & 0 & 0 & 0 & \sqrt{\frac{2 \xi_0^2}{t_c}} & 0 & 0 \\
        0 & 0 & 0 & 0 & 0 & 0 & 0 & 0 & \sqrt{\frac{1}{t_c'^2}} & 0 \\
        0 & 0 & 0 & 0 & 0 & 0 & 0 & 0 & 0 & \sqrt{\frac{1}{t_c'^2}} \\
    \end{pmatrix},
\end{equation}
\end{widetext}
and 
\begin{equation}
    \boldsymbol{\eta'} = (0\ 0\ \eta_{1x}\ \eta_{1y}\ 0\ 0\ \zeta_{x}\ \zeta_{y}\ \zeta_{x}'\ \zeta_{y}')^T.
\end{equation}
Introducing the correlation matrix $\boldsymbol{\Xi}$ with elements 
\begin{equation}
    [\Xi_{i,j}] = \langle \chi_i \chi_j \rangle - \langle \chi_i \rangle \langle \chi_j \rangle,
\end{equation}
it can be shown that $\boldsymbol{\Xi}$ satisfy the following equation:
\begin{equation}
    A\cdot \Xi + \Xi \cdot A^T + B\ B^T = 0.
    \label{eq:Xi_relation}
\end{equation}
Once the matrix relation Eq.~\eqref{eq:Xi_relation} is solved, the magnitude of the magnetic moment Eq.~\eqref{eq:M_def} can be calculated as
\begin{align}
    \langle M \rangle& = 
    \lim_{t\to \infty} \frac{|q|}{2} (\langle v_yx \rangle - \langle v_x y \rangle). \\ 
    &= \frac{|q|}{2} (\Xi_{4,1} - \Xi_{3,2}).
\end{align}
Substituting the values of $\Xi_{4,1}$ and $\Xi_{3,2}$, we obtain the average orbital magnetic moment as reported in Eq.~\eqref{eq:M_gen}.

\end{document}